\documentclass[%
aip,
 amsmath,amssymb,
 reprint,%
]{revtex4-1}

\usepackage{graphicx}
\usepackage{dcolumn}
\usepackage{bm}

\usepackage[utf8]{inputenc}
\usepackage[T1]{fontenc}
\usepackage{mathptmx}
\usepackage{etoolbox}

\newcommand{\be}{\begin{equation}}
\newcommand{\ee}{\end{equation}}

\newcommand{\vf}{\varphi}
\newcommand{\lf}{\left}
\newcommand{\rg}{\right}
\newcommand{\ra}{\rangle}
\newcommand{\la}{\langle}

\newcommand{\bea}{\begin{eqnarray}}
\newcommand{\eea}{\end{eqnarray}}
\renewcommand{\o}{\omega}

\newcommand{\nn}{\nonumber}

\usepackage{xcolor}
\usepackage{amsmath}
\usepackage{amssymb}
\usepackage{subfigure}
\usepackage{graphicx}
\usepackage{ulem}
\usepackage{multirow}
\usepackage{enumitem}
\usepackage{cases}
\usepackage{cases}
\usepackage{mhsetup}
\usepackage{mathtools}
\usepackage{amsbsy}
\usepackage{hyperref}
\usepackage{braket}
\makeatletter
\def\@email#1#2{%
 \endgroup
 \patchcmd{\titleblock@produce}
  {\frontmatter@RRAPformat}
  {\frontmatter@RRAPformat{\produce@RRAP{*#1\href{mailto:#2}{#2}}}\frontmatter@RRAPformat}
  {}{}
}%
\makeatother
\begin{document}

\preprint{AIP/123-QED}

\title{Flux-Tunable Regimes and Supersymmetry in Twisted Cuprate Heterostructures}
\author{Alessandro Coppo}
\affiliation{Istituto dei Sistemi Complessi, Consiglio Nazionale delle Ricerche, Via dei Taurini, 19 I-00185 Roma (IT)}
\affiliation{Physics Department,  University of Rome, ``La Sapienza'', P.le A. Moro, 2 I-00185 Roma (IT)}
\author{Luca Chirolli}
\affiliation{Quantum Research Center, Technology Innovation Institute, P.O. Box 9639 Abu Dhabi (UAE)}
\author{Nicola Poccia}
\affiliation{Leibniz Institute for Solid State and Materials Science Dresden (IFW Dresden), 01069 Dresden (DE)} 
\affiliation{Department of Physics, University of Naples Federico II, Via Cintia, Naples 80126 (IT)}
\author{Uri Vool}
\affiliation{Max Planck Institute for Chemical Physics of Solids, 01187 Dresden (DE)}
\author{Valentina Brosco}
\email[]{valentina.brosco@cnr.it}
\affiliation{Istituto dei Sistemi Complessi, Consiglio Nazionale delle Ricerche, Via dei Taurini, 19 I-00185 Roma (IT)}
\affiliation{Physics Department,  University of Rome, ``La Sapienza'', P.le A. Moro, 2 I-00185 Roma (IT)}
\date{\today}
\begin{abstract} 
Van der Waals (vdW) assembly allows for the creation of Josephson junctions in an atomically sharp interface between two exfoliated  Bi$_2$Sr$_2$CaCu$_2$O$_{8+\delta}$ (Bi-2212) flakes that are twisted relative to each other. In a narrow range of angles close to $45^\circ$, the junction exhibits a regime where time-reversal symmetry can be spontaneously broken and it can be used to encode an inherently protected qubit called flowermon.
In this work we investigate the physics emerging when two such junctions are integrated in a SQuID circuit threaded by a magnetic flux. 
We show that the flowermon qubit  regime is maintained up to a finite critical value of the magnetic field and, under appropriate conditions, it is protected against both charge and flux noise.  For larger external fluxes, the interplay between the inherent twisted d-wave nature of the order parameter and the external magnetic flux enables the implementation of different artificial atoms, including a flux-biased protected qubit and a supersymmetric quantum circuit.
\end{abstract}

\maketitle

Van der Waals (vdW) heterostructures, realized through the exfoliation and assembly of single atomic layers, are artificial quantum materials having widely tunable electronic and optical properties. Within these structures, the interplay of topological effects, strong correlation, and confinement can be precisely controlled by adjusting the interlayer twist angle yielding a wealth of  interesting phenomena including unconventional superconductivity,~\cite{cao2018}  topological ferromagnetic order,~\cite{sharpe2019} and correlated insulating states~\cite{cao2018a} just to mention a few.
 
Recently, the development of innovative  fabrication techniques allowed the isolation of atomically thin Bi$_2$Sr$_2$CaCu$_2$O$_{8+\delta}$ (Bi-2212) crystals~\cite{yu2019high, zhao2019sign, poccia2020spatially} with near-perfect superconductivity and lattice structure, and paved the way to the realization~\cite{zhao2023,Lee2023Encapsulating, martini2023twisted} of vdW heterostructures between twisted cuprate layers showing a strong dependence of the Josephson energy on the twist angle. The applied stacking technologies freeze the chemistry of the cuprate crystals below 200 K (-73 $^{\circ}$C) in ultra-pure argon atmosphere, and preserve the intrinsic and spatially-competing striped orders made of oxygen interstitials,~\cite{poccia2011evolution} incommensurate local lattice distortions and charge modulations~\cite{poccia2012optimum} as found in pristine cuprate single crystals.~\cite{fratini2010,campi2015inhomogeneity}

In these junctions, where detrimental disorder is reduced to a minimum, the d-wave nature of the superconducting state has significant effects on the junction characteristics. Particularly, within a narrow range of twist angles close to 45$^{\circ}$, it results in a strong suppression of single Cooper pair tunneling,~\cite{martini2023twisted} consequently making the contribution of two-Cooper-pair tunneling dominant.
In this regime, the Josephson energy has a leading $\cos(2\hat \vf)$ dependence on the superconducting phase difference $\hat \vf$ and the junction~\cite{can2021} hosts a peculiar superconducting state  where time-reversal symmetry can be spontaneously broken, in agreement with the experimental work in Ref.~\onlinecite{zhao2023}.
Further interesting topological phases were predicted at lower twist angles~\cite{lu2022,belanger2024} away from optimal doping or in more complex trilayer structures.~\cite{tummuru2022}

Very recently, it was proposed to utilize such twisted vdW cuprate junctions to realize novel  superconducting quantum devices.~\cite{brosco2024,patel2024} Specifically, the circuit design proposed in Ref.~\onlinecite{brosco2024}, nicknamed 'flowermon',  consists of a single vdW cuprate junction with a twist angle $\theta$ in the range $42^\circ - 48^\circ$,  shunted by a capacitor and coupled to a control and readout resonator in a circuit QED architecture.~\cite{Devoret_2013} The flowermon exploits the peculiar $\cos(2\hat \vf)$  dependence of the Josephson energy, stemming from the twisted d-wave nature of the order parameter, to encode a qubit inherently protected against capacitive noise. 

Capacitive fluctuations, arising from charge noise or stray electric fields, stand out as one of the most critical sources of noise limiting the coherence of many currently used superconducting qubits such as the transmon.~\cite{koch2007}
Over the years, significant research efforts focused on understanding  and characterizing  dielectric properties of materials~\cite{martinis_decoherence_2005,wang_surface_2015,murray2021} to reduce capacitive losses as well as on the development of alternative qubit designs  exploiting external magnetic fluxes or gates to drive the qubit to regimes with vanishing sensitivity to this kind of noise. Notable examples are the rhombus chain,~\cite{blatter_design_2001, doucot_topological_2003,gladchenko_superconducting_2009,bell_protected_2014} the $0-\pi$,~\cite{brooks_protected_2013,groszkowski_coherence_2018,gyenis_experimental_2021} the  bifluxon,~\cite{kalashnikov_bifluxon_2020} the blochnium,~\cite{pechenezhskiy2020} the KITE,~\cite{smith2020,smith_magnifying_2022} and semiconductor-superconductor~\cite{larsen_parity-protected_2020,schrade_protected_2022,ciaccia_charge-4e_2023,chirolli2024cooper} qubits, that can be employed also for hybrid topological protection schemes.~\cite{chirolli2021enhanced,chirolli2022swap}
A crucial difference between these qubits and the flowermon is that in the latter protection originates from the d-wave nature of the order parameter while in the former it is achieved through circuit engineering.  
In this regard, the concept of flowermon  is closely linked  to the pioneering theoretical ~\cite{blatter_design_2001,Ioffe1999,blais2000} and experimental~\cite{Bauch2006} research, which first explored the suppression of tunneling in d-wave based Josephson junctions to realize superconducting qubits. 

\begin{figure}
\includegraphics[width=\columnwidth]{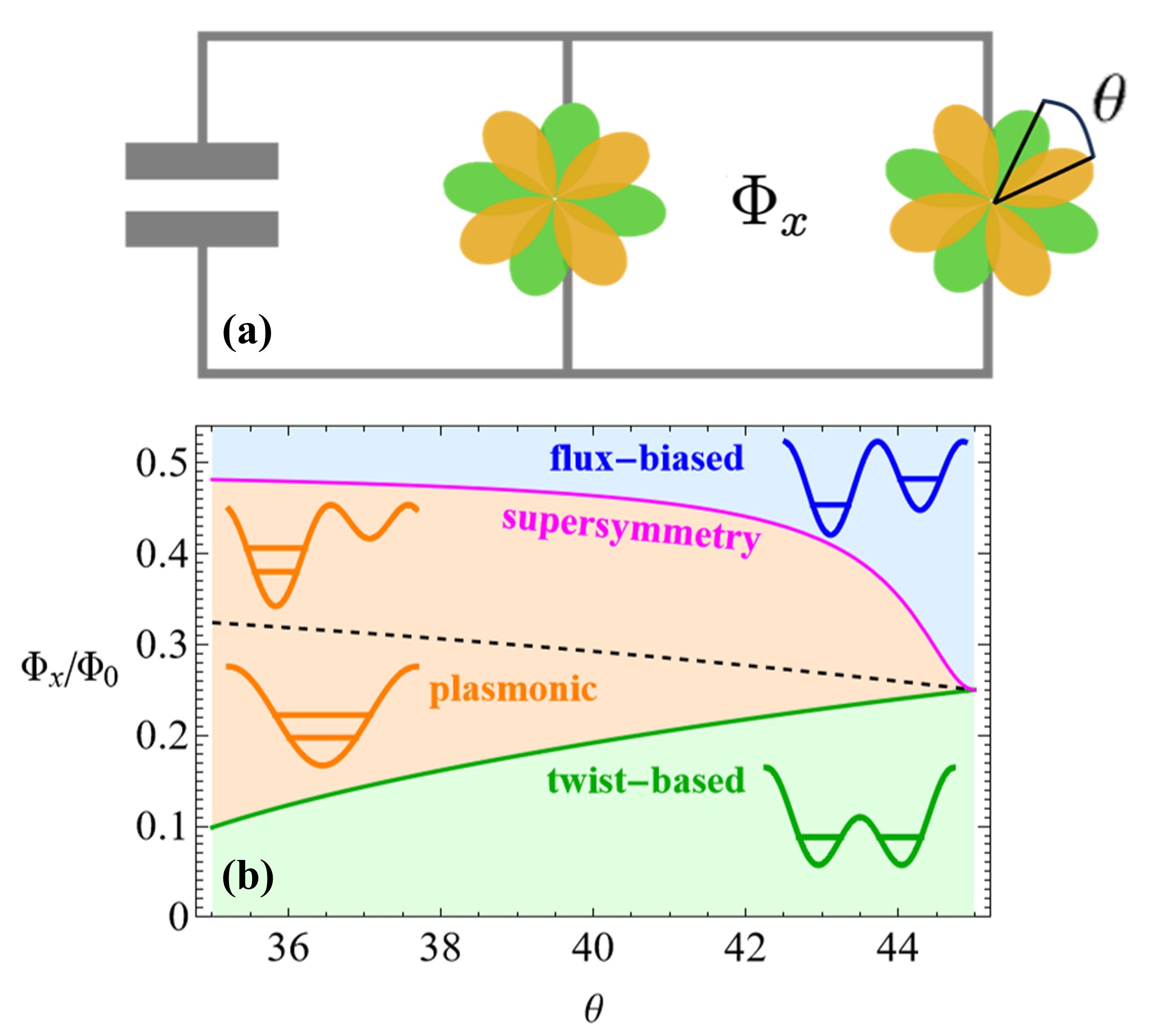}
  \caption{\textbf{(a)} Circuit scheme of the split-flowermon featuring two twisted BSCCO junctions  in a SQuID loop threaded by an external magnetic flux. \textbf{(b)} Scheme illustrating the different regimes as a function of the twist angle, $\theta$ and the external flux $\Phi_x$.}
    \label{circuit}
\end{figure}

In this Letter, we present a novel quantum device illustrated in Fig.~\ref{circuit}(a), comprising two twisted cuprate junctions  integrated in a Superconducting Quantum Interference Device (SQuID) loop and threaded by an external magnetic flux. By adjusting the external flux and the twist angle, this device can be tuned into various regimes hosting respectively: a symmetric, "twist-based", double-well potential, a ``plasmonic''  potential, and a ``flux-biased'' double-well potential, as illustrated in Fig.~\ref{circuit}(b). 
Below, we show how the structure of the low-energy spectrum changes across the different regimes leading to distinct sensitivities to charge and flux noise fluctuations.
The high-tunability of the device also enables the realization of a supersymmetric Hamiltonian where the spectrum has one non-degenerate ground-state and all other states are degenerate in pairs.
Supersymmetric spectra arise in superconducting circuits due to the non-trivial interplay of different tunneling mechanisms, such as rhombus elements,~\cite{ulrich2015} Majorana quasi-particles,~\cite{ulrich2014} or the charging energy spectrum of a single junction.~\cite{peyruchat2024} 
Here we show that supersymmetry marks the transition between the plasmonic and the flux-biased regimes, triggering significant changes in the coherence properties of the circuit.

To derive the circuit's  Hamiltonian, we begin by expressing the Josephson potential of a single vdW cuprate Josephson junction with an interlayer twist angle $\theta$ as the sum of the first and second harmonic Josephson tunneling as follows
\be \label{two-harm}
-E_{J\theta} \cos{\hat\vf}+E_{\kappa}\cos{2\hat\vf}~.
\ee
Here, $\hat\vf$ indicates the superconducting phase difference across the junction and higher-order harmonics are neglected assuming a weak tunnel coupling between the two flakes. In the above equation, $E_{J\theta}$ quantifies the energy associated with the tunneling of one Cooper pair across the junction. As discussed  in Refs.~\onlinecite{can2021,tummuru2022a} and confirmed experimentally in Ref.~\onlinecite{zhao2023}, $E_{J\theta}$ exhibits  a strong dependence on $\theta$:
\be E_{J\theta}=E_{J}\cos(2\theta)~.\label{ejtheta}\ee
Additionally,  $E_{\kappa}$ quantifies the energy associated with two Cooper pair tunneling. This term does not vanish  at $\theta=45^\circ$ and is predicted to have a weaker dependence on the twist angle,~\cite{can2021} that we neglect for simplicity. 

When two twisted Josephson junctions are integrated in a SQuID loop threaded by an external flux, $\Phi_x$, the Josephson's potential features four terms
\begin{align}\label{e. potential f-SQUID}
&U_{\phi_x}(\hat{\varphi})=-\big[E_{J_1\theta_1}\cos(\hat{\varphi}-\phi_x)+E_{J_2\theta_2}\cos(\hat{\varphi}+\phi_x)\big] \nonumber \\
&+\big[E_{\kappa_1}\cos{\left(2(\hat{\varphi}
-\phi_x)\right)}+E_{\kappa_2}\cos{\left(2(\hat{\varphi}+\phi_x)\right)}\big],
\end{align}
where  $E_{J_i\theta_i}$ and $E_{\kappa_i}$ with $i=1,2$ quantifying the tunnelling amplitudes of junctions 1 and 2, respectively, and $\phi_x=\pi\Phi_x/\Phi_0$ denotes the normalized flux with $\Phi_0=h/2e$ indicating the flux quantum. Note that $E_{J_1\theta_1}$ and $E_{J_2\theta_2}$ depend on the corresponding twist angles as dictated by Eq.~\eqref{ejtheta}, {\sl i.e.}
\begin{equation}
    E_{J_1\theta_1}=E_{J_1}\cos{(2\theta_1)},\quad E_{J_2\theta_2}=E_{J_2}\cos{(2\theta_2)}~.
\end{equation}
In Eq.~\eqref{e. potential f-SQUID}, the phase $\hat \vf$ denotes a symmetric gauge choice between the phases of the two junctions. Utilizing Eq.~\eqref{e. potential f-SQUID}, the whole circuit's Hamiltonian can be cast as
\begin{equation}\label{e. Hamiltonian f-SQUID}
\hat H = 4E_C\,(\hat{n}-\delta n_g(t))^2+U_{\phi_x}(\hat{\varphi})~,
\end{equation}
where the charging energy reads $E_C=e^2/(2C)$, $C$ being the shunting capacitance (see Fig.~\ref{circuit}(a)) which dominates over the internal capacitances,  $\hat n$ indicates the charge conjugate to $\hat\vf$ and $\delta n_g(t)$ accounts for charge fluctuations induced by external electric fields. Throughout this work we assume that the capacitance is sufficiently large that  the charging energy satisfies the relation $E_C\ll E_J,\, E_\kappa$, this condition defines the transmonic regime. Furthermore, we set $E_\kappa/ E_J=0.1$ following the prediction of Ref.~\onlinecite{can2021}. As discussed in more detail in Ref.~\onlinecite{brosco2024}, the value of $E_\kappa/E_J$ is crucially relevant to observe the double-well structure for a wide range of angles. Eventually, we set $\la \delta n_g\ra=0$: though our results should remain valid for the low energy levels of the spectrum regardless of charge bias, in this case, in the idealized situation of identical junctions, the circuit's Hamiltonian possesses various symmetries that enhance qubit coherence and simplify the analysis of the spectrum.~\cite{supplementary}

To explore both this idealized scenario and the more realistic case of small junctions asymmetry, it is advantageous to introduce the average twist angle, $2\theta=\theta_2+\theta_1$,  and the total tunneling energies, $E_{J}=E_{J_2}+E_{J_1}$ and  $E_{\kappa}=E_{\kappa_2}+E_{\kappa_1}$. Recasting the Josephson potential accordingly yields:
\bea\label{e. potential f-SQUID_2}
\!\!\!\!U_{\phi_x}(\hat{\varphi}) &\simeq&
\!\!\!- \widetilde E_J\cos(\hat{\varphi}-\vf_0)+\widetilde E_{\kappa}\cos(2(\hat \varphi-\vf_{0\kappa}))+\nn\\
& & \!\!\!\!\!\!\!\!\!\!\!\!-~2E_J d_\theta\sin{2\theta}\big[\sin{\phi_x}\sin{\hat{\varphi}}- d\cos{\phi_x}\cos{\hat{\varphi}}\big]
\eea
valid up to second-order corrections in the twist angle asymmetry $d_{\theta}=(\theta_2-\theta_1)/2$. In the above equation we introduced the effective tunneling energies,
\be
\widetilde{E}_J=E_J \cos(2\theta)\cos \phi_x \sqrt{1+d^2 \tan^2 \phi_x}
\ee
and 
\be
\widetilde{E}_{\kappa}=E_\kappa \cos 2\phi_x \sqrt{1+d_\kappa^2 \tan^2 2\phi_x}
\ee
where we denoted the asymmetry between the junctions as  
$d=(E_{J_2}-E_{J_1})/E_J$ and $d_{\kappa}=(E_{\kappa_2}-E_{\kappa_1})/E_\kappa$, and the angles $\vf_0$ and $\vf_{0\kappa}$ are defined by the following equations:
$\tan \vf_0=-d \tan \phi_x$ and $\tan 2\vf_{0\kappa}=-d_{\kappa} \tan 2\phi_x$.

\begin{figure}
 \hspace*{-2mm}\includegraphics[scale=0.42]{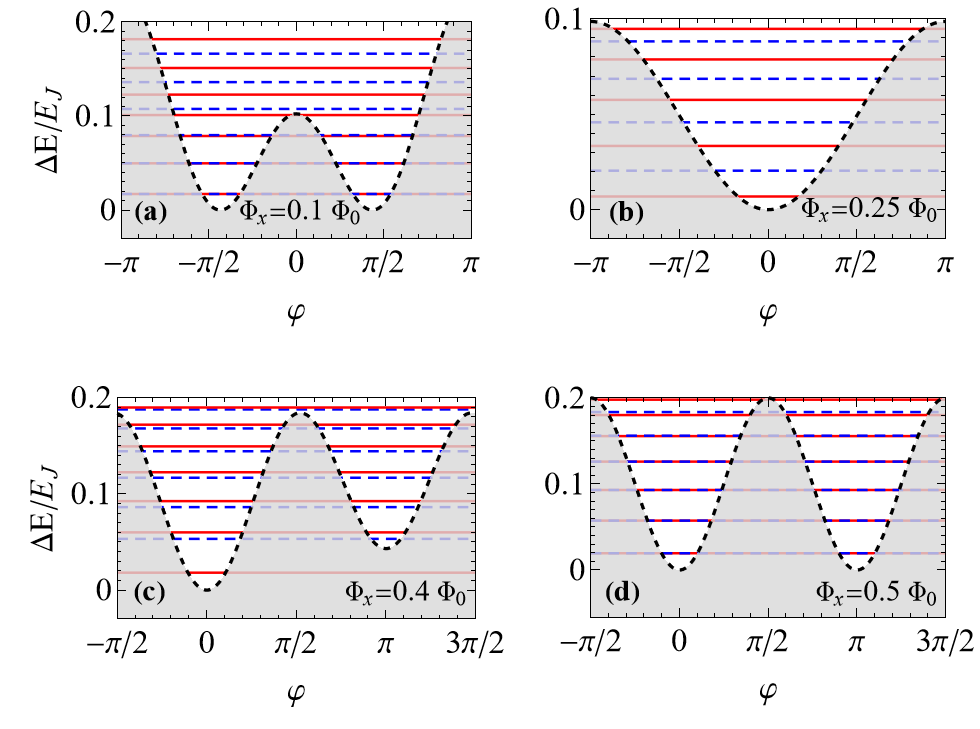}
  \caption{\textbf{Symmetrical flowermon SQuID potential and energy levels.} Shape of the potential for $\alpha=0.1$, $\theta=43^{\circ}$ and \textbf{(a)} $\Phi_x=0.1\;\Phi_0$, \textbf{(b)} $\Phi_x=0.25\;\Phi_0$, \textbf{(c)} $\Phi_x=0.4\;\Phi_0$ and \textbf{(d)} $\Phi_x=0.5\;\Phi_0$  together with the corresponding energy levels for $E_J/E_C=2000$.}
    \label{f.potential and levels}
\end{figure}

To keep the discussion simple, we first consider the case of identical junctions: $d=d_\kappa=0$, $d_\theta=0$. In this idealized  situation the external flux  controls the strength and sign of the effective Josephson tunnelings, $\widetilde E_J$, $\widetilde E_{\kappa}$, and,  depending on the twist angle $\theta$ and on the ratio  $\alpha=E_\kappa/E_J$,  it can be used to tune the Josephson potential.
Specifically, setting 
\be Y=\frac{4\alpha\cos(2\phi_x)}{\cos(2\theta)\cos(\phi_x)}\ee
 and assuming $\phi_x\in [0,\pi/2]$, we obtain that~\cite{supplementary}
(i) for $Y > 1$, the potential exhibits a symmetric double-well structure with minima at $\varphi = \pm \arccos(1/Y)$ as displayed in Fig.~\ref{f.potential and levels}(a);
(ii) for $|Y| < 1$, the potential features a single minimum at $\varphi = 0$, as shown in Fig.~\ref{f.potential and levels}(b) and 
(iii) for $Y < -1$, it displays an asymmetric double-well structure with minima at $0$ and $\pi$, as illustrated in Figs~\ref{f.potential and levels}(c)-(d).
\begin{figure}
  \hspace*{-4mm}\includegraphics[scale=0.29]{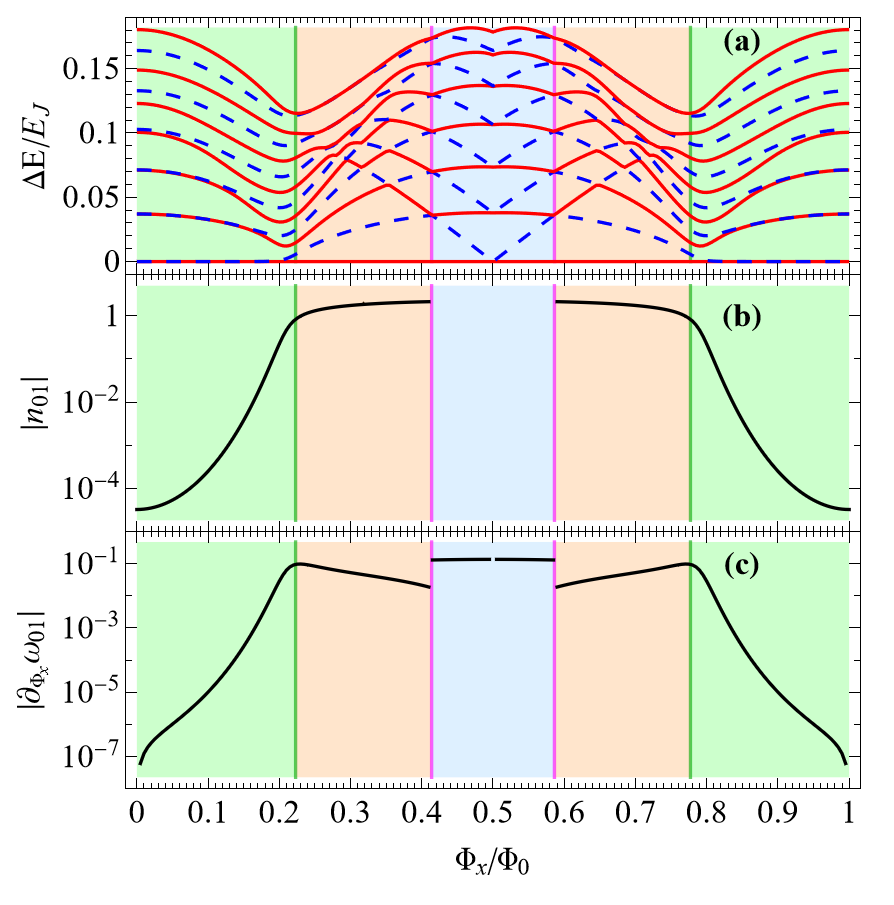}  
  \caption{\textbf{Energy spectrum, flux dephasing rate and charge relaxation rate} \textbf{(a)} Low-energy spectrum as function of $\Phi_x$. \textbf{(b)} Matrix element $|n_{01}|$ governing dielectric relaxation rate. In the central region it vanishes due to symmetry reason. \textbf{(c)}  Absolute value $|\partial_{\Phi_x}\omega_{01}|$ governing flux-dephasing normalized to $E_J/\Phi_0$. In all panels  background colors highlight the regimes of Fig.~\ref{circuit}(b).}
    \label{f.spectrum, flux dephasing and charge relax}
\end{figure}
As a result,  the parameter $Y$ controls  the symmetry of the low-energy eigenstates  and the structure of the spectrum. This dependence  gives rise to the regimes introduced in Fig.~\ref{circuit}(b) and underpins 
significant changes in  the system dynamics and susceptibility to external fluctuations.

The ``twist-based'' regime, realized at small external fluxes and high twist angles,  corresponds to $Y>1$ and it is characterized by a robust quasi-degeneracy of the low-energy levels as shown in Fig.~\ref{f.spectrum, flux dephasing and charge relax}(a).
In this regime two Cooper pair tunneling processes dominate the Josephson energy and the ground and first excited state have a well-defined Cooper pair number parity. 
In particular, analogous to what occurs in the flowermon,~\cite{brosco2024} the ground-state, $|\psi_0\ra$, contains  mostly even Cooper pair number states while  the first excited state, $|\psi_1\ra$, contains mostly odd Cooper-pair-number states. 
$|\psi_0\ra$  and  $|\psi_1\ra$  thus have very small overlap in the charge basis, yielding (see  green shaded area in Fig.~\ref{f.spectrum, flux dephasing and charge relax}(b)) an exponential suppression of the matrix element $n_{01}=\la\psi_0|\hat n|\psi_1\ra$. This leads to the exponential suppression of the relaxation rate induced by capacitive losses, $\Gamma_{1,{n_g}}$, that can be estimated as
\be \Gamma_{1,{n_g}}=\frac{(8 E_C)^2}{\hbar^2}\,S_{n_g}(\omega_{01})\,|n_{01}|^2    \ee
where $\hbar \omega_{01}$ is the  qubit energy  splitting and  $S_{n_g}(\omega)$ the spectral density of the capacitive noise. 
\begin{figure*}
   \hspace*{-5mm}\includegraphics[width=1.05\textwidth]{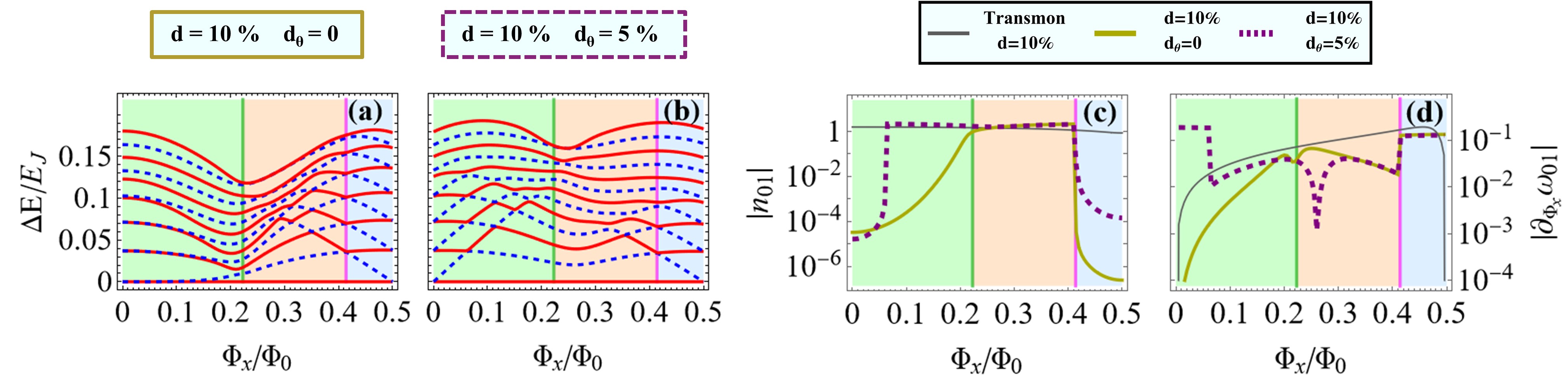}
  \caption{\textbf{The role of the asymmetries between the junctions.} \textbf{(a)} Low-energy spectrum as a function of $\Phi_x$ for $\alpha=0.1$, $\theta=43^\circ$, $d=10\%$ and $d_\theta=0$. \textbf{(b)} Low-energy spectrum as a function of $\Phi_x$  with $d_\theta=5\%$ and other parameters as in panel (a). \textbf{(c)} Matrix element of the charge operator governing dielectric relaxation between the two qubit states as a function of $\Phi_x$. \textbf{(d)} Absolute value of $|\partial_{\Phi_x}\omega_{01}|$ governing flux induced dephasing normalized to $E_J/\Phi_0$. In all panels, lines and background colors indicate the regimes for $d=0$, as in Fig.~\ref{circuit}(b).}
    \label{f.flux relax, dephasing and charge relax with asymmetry}
\end{figure*}
Furthermore, since the flux does not substantially affect the quasi-degeneracy of the levels, flux-noise induced dephasing, 
 \be  \Gamma_{\vf,{\Phi_x}}=
 \,S_{\Phi_x}(0)\,|\partial_{\Phi_x} \omega_{01}|^2~,\ee 
where $S_{\Phi_x}(\omega)$ is the spectral density of the flux-noise, is also exponentially suppressed in this region,~\cite{supplementary} as illustrated by the plot of  $|\partial_{\Phi_x} \omega_{01}|$ shown in  Fig.~\ref{f.spectrum, flux dephasing and charge relax}(c). 

As the external flux increases above a threshold value corresponding to $Y=1$, indicated by the green curve in Fig.~\ref{circuit}(b), the  system enters the ``plasmonic'' regime.~\cite{supplementary} 
In this regime, the low-energy eigenstates are confined within a single-well centered around $\vf=0$ and the charge relaxation matrix element becomes finite analogous to what happens in the transmon.~\cite{koch2007}
At even higher flux values, when $Y<-1$ (dashed black curve in Fig.~\ref{circuit}(b)) a second local minimum centered around $\vf=\pi$ appears in the potential. As long as this minimum is at high-energy, the structure of the low-energy spectrum does not change.
Correspondingly, the system remains in the plasmonic regime that thus includes the whole orange-shaded region in Fig.~\ref{circuit}(b). 

The important change in the dynamics of the system happens when the external flux crosses a threshold value in which the second minimum at $\vf=\pi$ begins to significantly affect the lowest excitation of the qubit (see Fig.~\ref{f.potential and levels}(c)). This threshold flux,
\be
\Phi_x^{\scalebox{0.7}{\rm SUSY}}=\frac{\Phi_0}{2\pi}\arccos\lf(\frac{-1}{1+16\alpha E_C/(E_J\cos^2(2\theta))}\rg)~,
\ee
is of special significance as it corresponds to a supersymmetry point of the Hamiltonian.
At this point, the spectrum decomposes into two decoupled subspaces having opposite symmetries under the parity operator $\hat K$ which satisfies $\hat K\hat{\varphi}\hat{K}=-\hat{\varphi}$ and implements the time-reversal symmetry for $\Phi_x=0$.
The ground-state is $\hat K$-symmetric and  non-degenerate while all excited levels feature a pair of degenerate states with opposite symmetry.
Specifically, the two lowest excited states, with wavefunctions corresponding to a plasmonic first excited state centered in the $\vf=0$ well, and a Gaussian-like state centered at the $\vf=\pi$ well, are exactly degenerate at this point.

For flux values above the supersymmetry point, $\Phi_x>\Phi_x^{\scalebox{0.7}{\rm SUSY}}$, the ground state remains a Guassian around the $\vf=0$ well while the first excited state becomes a Gaussian centered at the $\vf=\pi$ well. 
This marks the flux-biased region, in which the ground and first excited states are both symmetric under the parity operator $\hat K$ and the matrix element $n_{01}$ vanishes exactly. For similar reasons, however, the flux-derivative of the qubit frequency becomes finite, yielding a finite $\Gamma_{\vf,{\Phi_x}}$ as shown in Fig.~\ref{f.spectrum, flux dephasing and charge relax}(c). 
This problem is general to all flux-biased $\cos(2\hat \vf)$ qubits shunted by a large capacitor, as thoroughly illustrated in Ref.~\onlinecite{dodge2023}.

It is worth noticing that in all regimes, the dephasing induced by charge noise and the relaxation induced by flux-noise are  suppressed by the large shunt capacitance  and symmetry considerations,~\cite{supplementary} and therefore in the manuscript we focused on charge-induced relaxation and flux-induced dephasing, which are the dominant loss mechanisms.

We now consider the role of the asymmetries between the junctions described by the parameters $d$, $d_\kappa$ and $d_\theta$ according to Eq.~\eqref{e. potential f-SQUID_2}. 
Since the asymmetries in the tunneling energies are mostly due to geometric factors, we set $d_\kappa=d$. 
However, we separately consider the effect of the twist angle asymmetry $d_\theta$.  
Furthermore, as in the symmetric case, we focus on the matrix elements governing charge-induced relaxation and flux-induced dephasing since these are the dominant loss mechanisms in the asymmetric case as well, as described in the supplementary material. There are two interesting features to be underlined in the results shown in Fig.~\ref{f.flux relax, dephasing and charge relax with asymmetry}. First, we note that $d$ and $d_\kappa$ do not undermine the inherent protection against charge and flux noise characterizing the twist-based regime, especially compared to an asymmetric transmon, see Fig.~\ref{f.flux relax, dephasing and charge relax with asymmetry}(c)-(d). Furthermore, while introducing small $d=d_\kappa$ breaks the supersymmetry, it still preserves the quasi-degeneracy of the low energy states.
Second, $d_\theta$ is much more detrimental and it leads to a strong enhancement  of both charge-induced relaxation and flux-induced dephasing across a wide range of fluxes. The sharp features in the flux dependence of the flux-induced dephasing with finite $d_\theta$ shown in  Fig.~\ref{f.flux relax, dephasing and charge relax with asymmetry}(d) are not universal but related to the specific choice of the parameters.

In conclusion, we developed a novel device based on a SQuID loop of two junctions formed by twisted cuprate heterostructures. By manipulating the external flux in the loop, the device can be tuned to substantially different regimes and it features the interplay of different mechanisms of protection against decoherence. At low values  of the external flux the circuit  maintains the protection against charge noise offered by a single-junction flowermon with the added benefit of tunable energy levels. This protection can be traced back to the inherent d-wave nature of the junctions which preserves the  double-well structure and symmetry of the potential even in the presence of external flux. At flux values close to $\Phi_x=\Phi_0/2$, the circuit develops a double-well potential by a more conventional flux-biased mechanism. This regime also shows significant protection from charge noise, but dephasing due to flux fluctuations in the loop is not suppressed. The critical flux at which the circuit enters the flux-biased regime is a special point where the spectrum exhibits a supersymmetric structure.
We demonstrated that this point marks a change in the symmetry properties of the excited state and  yields sharp modifications in the system's coupling to external noise fluctuations and, consequently, in the decoherence rates. The role of imperfections in junction fabrication was also investigated leading to the discovery that the system is robust to  the asymmetry between the energy of the junctions but highly sensitive to the asymmetry in the twist angles. 
This circuit therefore  offers the opportunity to explore fundamental problems in quantum physics and to develop new devices for quantum technology.
Furthermore, it contributes to illustrate how integrating new materials and heterostructures into quantum superconducting circuits can unveil intriguing and novel phenomena opening new research pathways and triggering further progresses in fabrication technology.
The possibility to experimentally realize the  device proposed in the present work critically depends on the ability to fabricate high-quality twisted interfaces and to integrate them in quantum superconducting nanocircuits. A first advancement in this direction discussed in Ref.~\onlinecite{saggau2023,MartiniMIckey}, proposes to decouple the circuit fabrication from the twisted vdW junction fabrication process harnessing integration within a silicon nitride membrane.\\

Acknowledgements - 
V.B. and A.C. acknowledge financial support from PNRR MUR project PE0000023-NQSTI
financed by the European Union  – Next Generation EU.
The work is partially supported
by the Deutsche Forschungsgemeinschaft (DFG 512734967, DFG 492704387, and DFG 460444718), co-funded by the European Union (ERC-CoG, 3DCuT, 101124606 and ERC-StG, cQEDscope, 101075962). The authors are deeply grateful to Bernard van Heck, Kornelius Nielsch, Francesco Tafuri, Domenico Montemurro, Davide Massarotti, for support and illuminating discussions.

\bibliography{Biblio}


\end{document}


\title{Supplementary material for ``Flux-tunable regimes and supersymmetry in Twisted Cuprate Heterostructures''}
\author{Alessandro Coppo}
\affiliation{Istituto dei Sistemi Complessi, Consiglio Nazionale delle Ricerche, Via dei Taurini, 19 I-00185 Roma, Italy}
\affiliation{Physics Department,  University of Rome, ``La Sapienza'', P.le A. Moro, 2 I-00185 Roma, Italy}
\author{Luca Chirolli}
\affiliation{Quantum Research Center, Technology Innovation Institute, P.O. Box 9639 Abu Dhabi, UAE}
\author{Nicola Poccia}
\affiliation{Leibniz Institute for Solid State and Materials Science Dresden (IFW Dresden), 01069 Dresden, Germany} 
\affiliation{Department of Physics, University of Naples Federico II, Via Cintia, Naples 80126, Italy}
\author{Uri Vool}
\affiliation{Max Planck Institute for Chemical Physics of Solids, 01187 Dresden, Germany}
\author{Valentina Brosco}
\affiliation{Istituto dei Sistemi Complessi, Consiglio Nazionale delle Ricerche, Via dei Taurini, 19 I-00185 Roma, Italy}
\affiliation{Physics Department,  University of Rome, ``La Sapienza'', P.le A. Moro, 2 I-00185 Roma, Italy}
\date{\today}

{
\let\clearpage\relax
\maketitle
}
\section{The Josephson potential for identical junctions}
\noindent In this section we study the Josephson potential describing the device introduced in the main text, that is a SQuID comprising two twisted cuprate junctions. In particular, we here consider identical junctions and  we investigate the change in the potential shape when the parameter $Y$ (see Eq.~\eqref{e. U_2} below) is varied.
Let us start by  writing the  expression of the  potential as follows
\begin{equation}\label{e. potential symmetric f-SQUID}
U_{\phi_x}(\hat\varphi)=-\widetilde{E}_J\cos(\hat\varphi)+\widetilde{E}_\kappa\cos(2\hat\varphi)\quad\mbox{with}\quad\widetilde{E}_J=E_J\cos(2\theta)\cos(\phi_x)~,\quad\widetilde{E}_\kappa=E_\kappa\cos(2\phi_x)~.
\end{equation}
Here, as in the main text, $\hat\varphi$ denotes a symmetric gauge choice between the superconducting phases of the two junctions, $E_J$ and $E_\kappa$ quantify the energies associated with the one and two Cooper pairs tunneling, $\theta$ is the interlayer twist angle and $\phi_x/\pi$ the external flux $\Phi_x$ normalized with the flux quantum $\Phi_0$. In this section we analyze the analytical structure of $U_{\phi_x}(\hat\varphi)$. 
We start by noting that, since
\begin{equation}
    U_{\pi\pm\phi_x}(\varphi)= U_{\phi_x}(\varphi+\pi)~,
\end{equation}
we can consider $0\leq\phi_x\leq \pi/2$. For similar reasons we can set $0\leq\theta\leq 45^\circ$. Within these assumptions we have $\widetilde{E}_J>0$ and we can recast  Eq.~\eqref{e. potential symmetric f-SQUID} for $\theta\neq 45^\circ$ and $\phi_x\neq \pi/2$ as
\begin{equation}\label{e. U_2}
U(\varphi)=\frac{\widetilde{E}_J}{4}\big[-4\cos(\varphi)+Y\cos(2\varphi)\big]\qquad\mbox{where}\qquad Y=\frac{4\widetilde{E}_\kappa}{\widetilde{E}_J}=\frac{4\alpha\cos(2\phi_x)}{\cos(2\theta)\cos(\phi_x)}~.
\end{equation}
with $\alpha=E_\kappa/E_J$.
The cases $\theta=45^\circ$ and $\phi_x=\pi/2$ will be discussed in Sec.~\ref{0pi_vs_pihalf}. 
The first and second derivatives of $U(\varphi)$ can be then  expressed as
\begin{align}\label{e. potential f-SQUID_ddK0 der}
&U'(\varphi)=\widetilde{E}_J\,\sin(\varphi)\big[1-Y\cos(\varphi)\big]~,\\
&U''(\varphi)=\widetilde{E}_J\,\big[\cos(\varphi)-Y\cos(2\varphi)\big]~.
\end{align}
The above equations yield stationary points at $\varphi=0$, $\vf=\pi$ and,  for $|Y|>1$, also at $\varphi=\pm\varphi^*$ with $\varphi^*=\arccos{\left(1/Y\right)}$. Evaluating the corresponding second derivatives
\begin{align}
    &U''(0)=\widetilde{E}_J\,(1-Y)~,\\
    &U''(\pm\pi)=-\widetilde{E}_J\,(Y+1)~,\\
    &U''(\pm\varphi^*)=\widetilde{E}_J\,(Y^2-1)/Y~,
\end{align}
we easily show that:
\vspace*{-1mm}
\begin{itemize}
\item[(i)] for $Y>1$ it has minima at  $\varphi=\pm\varphi^*$ with $\varphi^*<\pi/2$ and maxima at $\varphi=0$ and $\vf=\pi$;
\vspace*{-2mm}
\item[(ii)] for $|Y|\leq1$ it has a minimum at $\vf=0$ and a maximum at $\vf=\pi$;
\vspace*{-2mm}
\item[(iii)]for $ Y<-1$ the potential has minima at $\varphi=0$ and $\vf=\pi$ and maxima at $\varphi=\pm\varphi^*$ with $\varphi^*>\pi/2$.
\end{itemize}

\begin{figure}[H]
   \centering
   \includegraphics[scale=0.65]{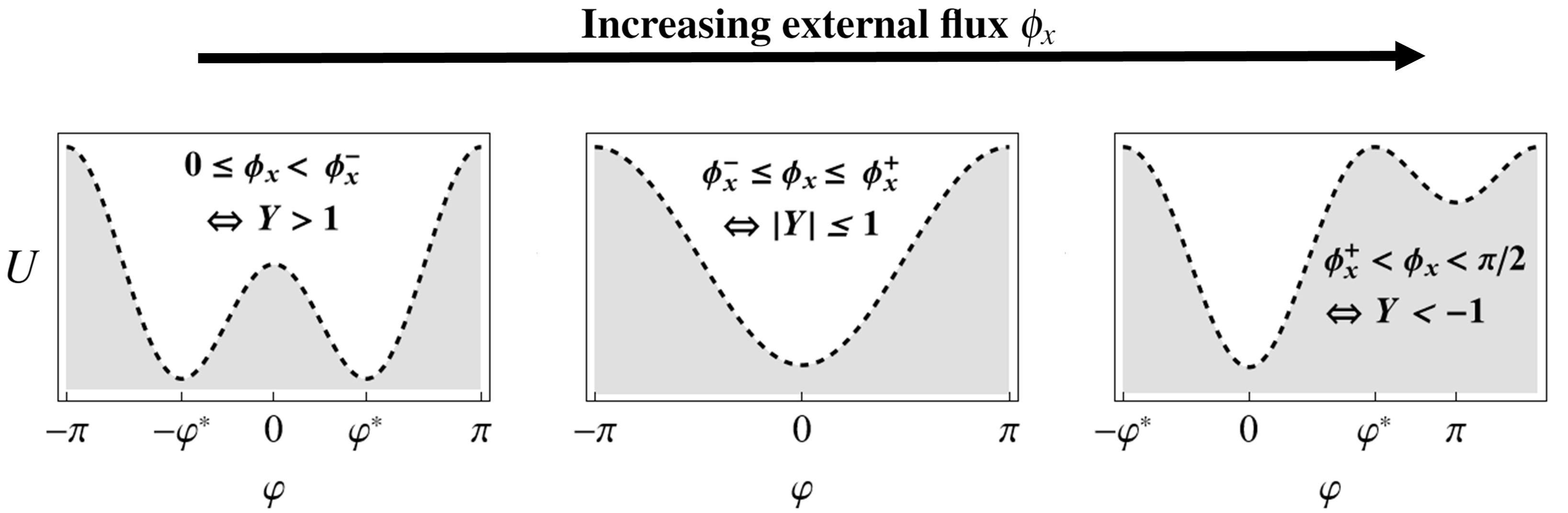}
  \caption{Shape of the Josephson potential for increasing external flux.}
    \label{FS1}
\end{figure}

\noindent We therefore introduce the critical flux values  $\phi_x^{\pm}$ such that $Y(\phi_x^{\pm})=\mp 1$ given by 
\begin{equation}
    \phi_x^{\pm}=\arccos\left(\frac{\sqrt{1+2 a^2}\mp 1}{2a}\right)\quad \mbox{with}\quad a=8\alpha/\cos(2\theta),
\end{equation}
satisfying $\phi_x^-<\pi/4<\phi_x^+$ and we distinguish three cases, as depicted in Fig.~\ref{FS1}. For  $0\leq\phi_x<\phi_x^-$ the potential has a symmetric double-well structure; for $\phi_x^-\leq\phi_x\leq\phi_x^+$ it is single-well and for $\phi_x^+<\phi_x<\pi/2$ it yields an asymmetric double-well. 
The solid green line and the dashed black line of Fig.~1(b) of main text represent respectively $\phi_x^-$ and $\phi_x^+$.  
Note that, while the asymmetric double-well regime only requires a sufficiently large flux, the symmetric double-well regime can be achieved  only for $\theta\geq\theta_c$, with
\begin{equation}
    \theta_c=
    \begin{dcases}
        \arccos{(4\alpha)}/2\quad\mbox{if}\quad \alpha<1/4~,\\
         0\qquad\qquad\qquad\;\mbox{if}\quad \alpha\geq 1/4~.
    \end{dcases}
\end{equation}
For this reason we refer to this regime as twist-based.

\section{Symmetry properties of the qubit eigenstates}\label{diff regimes}

\begin{figure}[H]
  \includegraphics[scale=0.65]{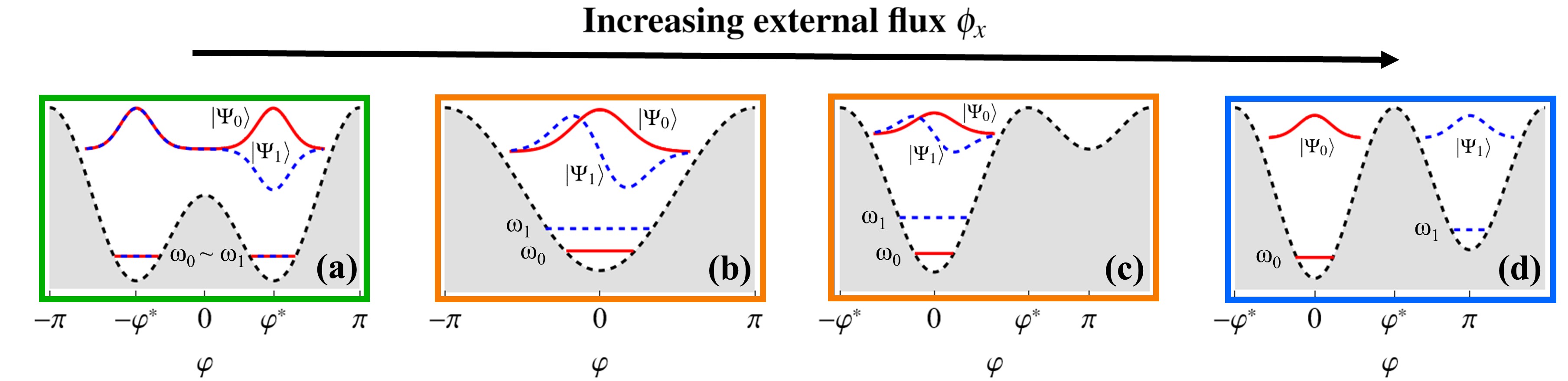}
  \caption{First two energy levels and wavefunctions for increasing external flux; the solid red lines identify $\omega_0$ and $\ket{\Psi_0}$, while the dashed blue ones $\omega_1$ and $\ket{\Psi_1}$. The plots frames are coloured according to the different regimes as in Fig.~1(b) of the main text.}
    \label{FS2}
\end{figure}

\noindent The circuit Hamiltonian can be cast as
\begin{equation}\label{e. Hamiltonian f-SQUID}
\hat H = 4E_C\,(\hat{n}-\delta n_g(t))^2+U_{\Phi_x}(\hat{\varphi})~,
\end{equation}
where $E_C$ is the charging energy,  $\hat n$ indicates the charge conjugate to $\hat\vf$ and $\delta n_g(t)$ accounts for charge fluctuations due to external electric fields. In this section we set the average $\langle\delta n_g\rangle$ at zero.  In order to discuss the relaxation and dephasing rates induced by charge and flux-noise, we focus on the first two energy levels $\hbar \omega_0$ and $\hbar \omega_1$  and the corresponding wavefunctions $\ket{\Psi_0}$ and $\ket{\Psi_1}$ shown in Fig.~\ref{FS2}.  We assume that the charging energy $E_C$  is sufficiently small to keep the two levels inside the well and below the barrier.
\noindent For increasing external flux the device enters the regimes introduced in the main text (see Fig.~1(b) of the main text), namely, the twist-based, the plasmonic and the flux biased regimes.
\begin{itemize}
    \item \textbf{Twist-based regime}\\
In the twist-based regime the qubit wavefunctions are localized inside a symmetric double-well with a barrier that decreases with increasing flux as shown in Fig.~\ref{FS3}(a) and have an exponentially small frequency splitting $\omega_{01}=\omega_1-\omega_0$ as shown in Fig.~\ref{FS3}(b). Since the Hamiltonian commutes with the parity operator $\hat{K}$, satisfying $\hat K\hat{\varphi}\hat{K}=-\hat{\varphi}$, non-degenerate states have definite parity w.r.t. $\varphi=0$. In particular $\ket{\Psi_0}$ is $\hat{K}$-symmetric and $\ket{\Psi_1}$ is $\hat{K}$-antisymmetric. A further consequence of the structure of the potential, having minima at $\vf =\pm \varphi^*\sim \pi/2$ is that the wavefunctions are  approximately $\pi$-periodic and $\pi$-antiperiodic .  This in turn implies that they have a well-defined Cooper pair number parity in the charge basis, {\sl i.e.} $\ket{\Psi_0}$ contains mostly even Cooper pair number states and $\ket{\Psi_1}$ odd Cooper pair number states. This is illustrated in Fig.~\ref{FS3}(c) where we plot the projection of  $\ket{\Psi_0}$ and $\ket{\Psi_1}$ on even Cooper pair number states, $\sum_n|\braket{2n|\Psi}|^2$, with $\ket{n}$ denoting the eigenstates of the charge operator $\hat{n}$.
    \begin{figure}[H]
    \centering
    \begin{minipage}{0.8\textwidth}\hspace*{-9mm}\includegraphics[scale=0.45]{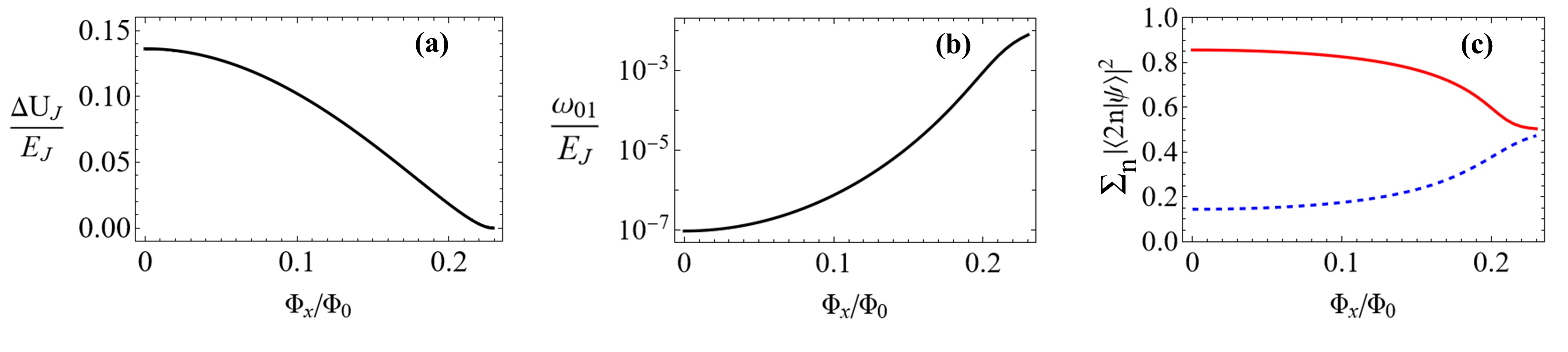}
    \end{minipage}
  \hspace*{8mm}\begin{minipage}{0.9\textwidth}
  \caption{\textbf{(a)} The height $\Delta U_J$ of the barrier, \textbf{(b)} the splitting $\omega_{01}$ and \textbf{(c)} the scalar product $\sum_n|\braket{2n|\Psi}|^2$  for the first two levels ($\ket{\Psi_0}$ red curve, $\ket{\Psi_1}$ blue curve) in the twist-based regime. On the horizontal axes the external flux increases up to $\phi_x=\phi_x^-$, where the double-well closes. The twist angle, the tunnelling energies and the charging one are as in the main text, i.e. $\theta=43^\circ$, $\alpha=0.1$ and $E_J/E_C=2000$ for which $\phi_x^-\sim 0.229\,\pi$.}
    \label{FS3}
    \end{minipage}
\end{figure}

\vspace*{-4mm}
\item \textbf{Plasmonic regime}\\
The regime depicted by Fig.~\ref{FS2}(b)-(c) is governed by a plasmonic potential with low-energy eigenstates confined within a single-well centered around the minimum at $\varphi=0$. The lowest-energy  eigenstates $\ket{\Psi_0}$ and $\ket{\Psi_1}$ are, as in the twist-based regime, $\hat{K}$-symmetric and $\hat{K}$-antisymmetric, but they have a finite energy splitting. For $\phi_x=\phi_x^+$ a second local minimum centered around $\varphi=\pi$ appears. This condition is indicated by the dashed black line in Fig.~1(b) of the main text. Nevertheless, the presence of this local high-energy minimum does not influence the properties of the low-energy spectrum as long as $\phi_x<\phi_x^{\scalebox{0.7}{\rm SUSY}}$.

\item \textbf{Flux-biased regime}\\
As the flux overcomes the threshold  value $\phi_x^{\scalebox{0.7}{\rm SUSY}}$ the system enters the flux-biased regime and the spectrum undergoes significant changes. Namely, the lowest excited state becomes $\hat{K}$-symmetric and its center moves from $\varphi=0$ to $\vf=\pi$. As stated in the main text, and demonstrated in Sec.~\ref{SUSY}, $\phi_x^{\scalebox{0.7}{\rm SUSY}}$ is a supersymmetry point. As shown in Fig.~\ref{FS4}(a), the splitting $\omega_{01}$ decreases with $\phi_x$ following the energy of the local potential minimum. Note that in the flux-biased regime, if $\phi_x$ is sufficiently close to $\pi/2$, the wavefunctions $|\Psi_0\ra$ and $|\Psi_1\ra$ are substantially equal up to a translation by $\pi$ as illustrated in Fig.~\ref{FS4}(b)-(c) where we plot the imbalance $\sigma_{01}=2(\sigma_1-\sigma_0)/(\sigma_1+\sigma_0)$ between their standard deviations $\sigma_0$, $\sigma_1$ and the scalar product $C_{01}=\int d\!\vf\, \psi_0(\vf)\, \psi_1(\vf-\pi)$. As a consequence these wavefunctions are thus equal on even Cooper pair number states and differ by a sign on odd ones. Moreover if $E_C\ll E_J$ they stay also localized around $\vf=0$ and $\vf=\pi$ as shown by  Fig.~\ref{FS4}(d).\\

 \begin{figure}[H]
    \centering
    \begin{minipage}{0.82\textwidth}\hspace*{-13mm}\includegraphics[scale=0.43]{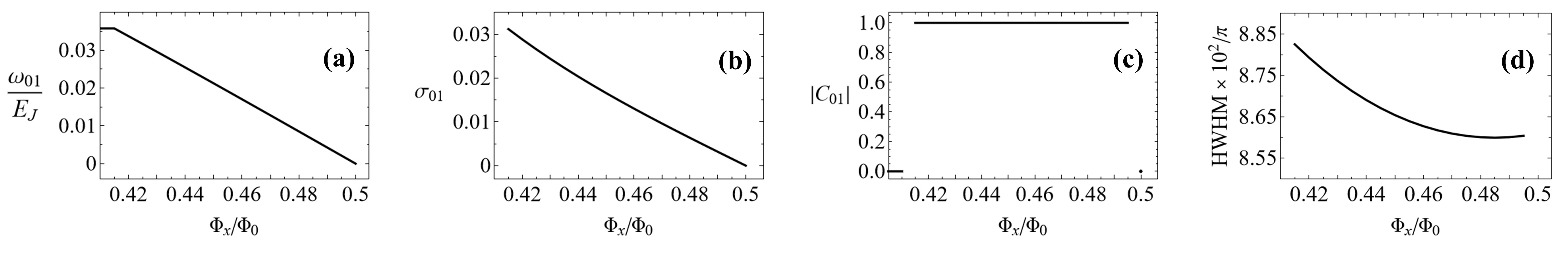}
    \end{minipage}
  \hspace*{8mm}\begin{minipage}{0.9\textwidth}
  \caption{\textbf{(a)} The splitting $\omega_{01}$, \textbf{(b)} the standard deviations imbalance $\sigma_{01}$, \textbf{(c)} the scalar product $C_{01}$ and \textbf{(d)} the half width half maximum (HWHM) for the first two levels in the flux-biased regime. On the horizontal axes the external flux increases from $\phi_x^{\scalebox{0.7}{\rm SUSY}}$ to $\pi/2$. All the panels are obtained for the parameters $\theta$, $\alpha$ and $E_J/E_C$ set as in Fig.~\ref{FS3} for which $\phi_x^{\scalebox{0.7}{\rm SUSY}}\sim 0.414\,\pi$.}
    \label{FS4}
    \end{minipage}
\end{figure}
Notice that, being $E_C$ sufficiently small to prevent  $\hbar \omega_0$ and $\hbar{\omega_1}$ from approaching the top of the well, we can obtain the plots in Fig.~\ref{FS4} by means of a Gaussian approximation. In particular, according to $\theta\sim 45^\circ$ and $\phi_x\sim\pi/2$, it is $\hbar\omega_{01}\sim E_J(4-\sqrt{2E_\kappa/E_C})(45^\circ-\theta)(\pi/2-\phi_x)$ {\sl i.e.} $\omega_{01}$ scales linearly with $\phi_x$.
\end{itemize}

\section{The rates induced by charge and flux-noise}
\noindent The relaxation and the dephasing rates $\Gamma_1$ and $\Gamma_\varphi$ induced by charge and flux-noise can be estimated within Fermi's golden rule as 
\begin{equation}
    \begin{dcases}
        \Gamma_{1,{n_g}}=\frac{(8 E_C)^2}{\hbar^2}\,S_{n_g}(\omega_{01})\,|n_{01}|^2~,\\
        \Gamma_{\varphi,{n_g}}=\frac{(8 E_C)^2}{\hbar^2}\,S_{n_g}(0)\,|n_{11}-n_{00}|^2~,
    \end{dcases}\quad\mbox{and}\quad
    \begin{dcases}
\Gamma_{1,{\Phi_x}}=\,S_{\Phi_x}(\omega_{01})\,|f_{01}|^2~,\\
    \Gamma_{\vf,{\Phi_x}}=\,S_{\Phi_x}(0)\,|f_{11}-f_{00}|^2~,
\end{dcases}
\end{equation}
with $S_{n_g}(\omega)$ and $S_{\Phi_x}(\omega)$ denoting the noise spectral densities and $n_{ij}=\braket{\Psi_i|\hat{n}|\Psi_j}$, $f_{ij}=\braket{\Psi_i|\partial_{\Phi_x}\hat{U}|\Psi_j}$ with $i,j\in\{0,1\}$.
The behavior of the rates in the different regimes is therefore strongly influenced by the matrix elements $f_{ij}$ and $n_{ij}$ that in turn depend on the symmetry and shape of the wavefunctions.
\begin{figure}[H]
   \centering
   \includegraphics[scale=0.45]{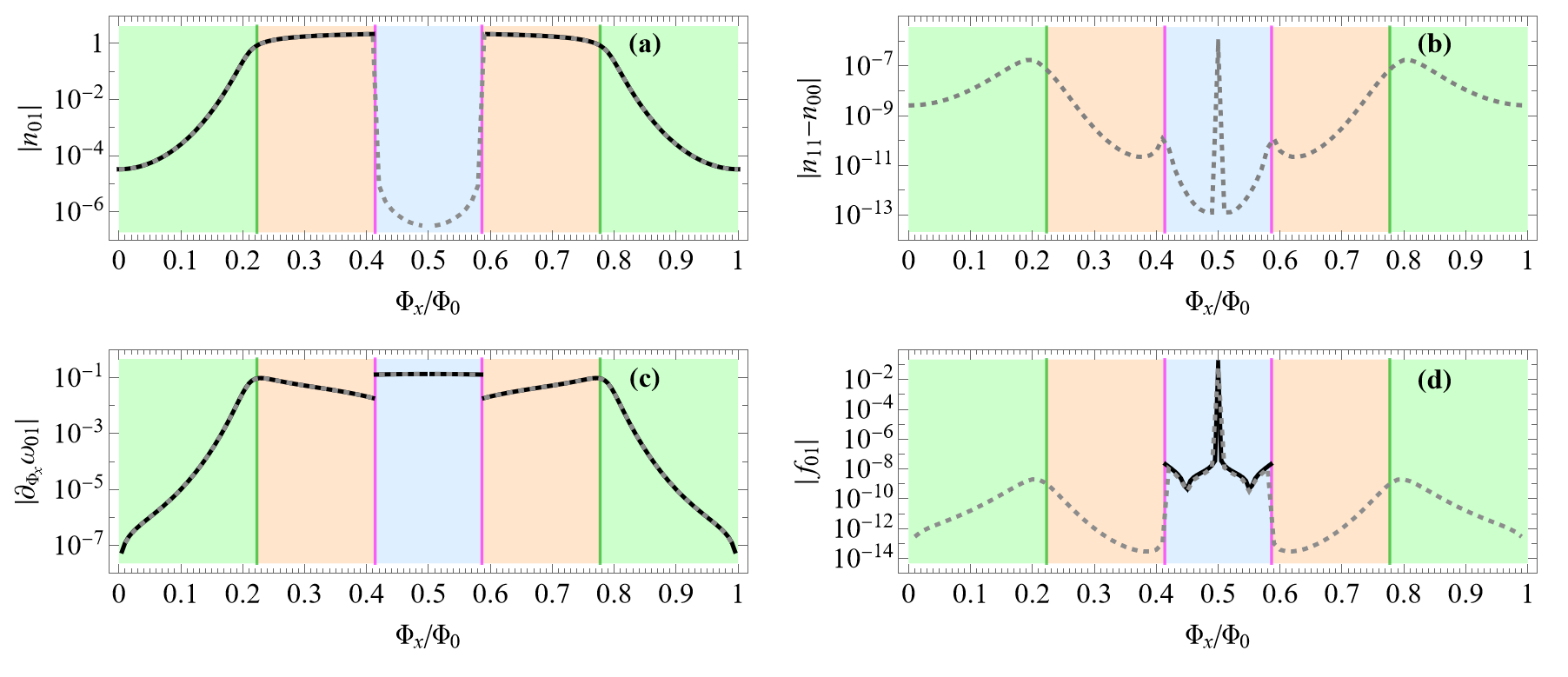}
  \caption{The matrix elements governing the rates induced by charge and flux-noise for $\langle n_g\rangle=0$ (solid black curves) and for $\langle n_g\rangle=0.25$ (dashed gray curves). In all panels, lines and background colors indicate the regimes for $\langle n_g\rangle=0$, as in Fig. 1(b) of the main text.}
    \label{FS5}
\end{figure}
\begin{itemize}

\item \textbf{Dielectric relaxation} $\Gamma_{1,{n_g}}\propto|n_{01}|$\\
\vspace*{-2mm}\\
In the twist-based regime $n_{01}$ is exponentially suppressed by the quasi-localization of $\ket{\Psi_0}$ and $\ket{\Psi_1}$ in the charge basis. In order to discuss the behavior of $n_{01}$ in the other regimes, let us remark that the charge operator $\hat n$ anticommutes with $\hat{K}$, $\{\hat{K},\hat{n}\}=0$, {\it i.e}. it is $\hat{K}$-antisymmetric.
Therefore since in the flux-biased regime the two wavefunctions are both $\hat{K}$-symmetric, $n_{01}$ is cancelled out by parity. On the other hand it stays significantly different from zero in the plasmonic regime.
If the average $\langle n_g\rangle$ does not vanish the symmetry of the wavefunctions changes but, as shown in Fig. \ref{FS5}(a), we still find an exponential suppression of $n_{01}$ in the flux-biased and twist-based regimes.

\item \textbf{Charge dephasing} $\Gamma_{\varphi,{n_g}}\propto|n_{11}-n_{00}|$\\
\vspace*{-2mm}\\
At $\langle n_g\rangle=0$ the matrix element $|n_{11}-n_{00}|$ vanishes in all the regimes since $|\braket{n|\Psi_0}|^2$ and $|\braket{n|\Psi_1}|^2$ are $\hat{K}$-symmetric. Away from $\langle n_g\rangle=0$ the protection against charge-induced dephasing is granted by $E_C\ll E_J, E_\kappa$ for which the dependence of the spectrum on $\langle n_g\rangle$ is exponentially suppressed in $E_J/E_C$. See Fig.~\ref{FS5}(b).

\item \textbf{Flux dephasing} $\Gamma_{\varphi,{\Phi_x}}\propto|f_{11}-f_{00}|$\\
\vspace*{-2mm}\\
According to the Hellmann–Feynman theorem, $f_{11}-f_{00}=\partial_{\Phi_x} (\hbar \omega_{01})$. Therefore, analogous to $\omega_{01}$, $f_{11}-f_{00}$ is exponentially suppressed in the twist-based regime and significantly different from zero in the other ones. Moreover, it has a jump at $\phi_x^{\scalebox{0.7}{\rm SUSY}}$ since $\omega_{01}$ has a cusp at this point. Finally, $|f_{11}-f_{00}|$ is approximately constant in the flux-biased regime since  $\omega_{01}$ is linear for $\phi_x^{\scalebox{0.7}{\rm SUSY}}\!\sim\pi/2$.  The behavior seems essentially the same also for $\langle n_g\rangle\neq0$ as shown Fig.~\ref{FS5}(c).

\item \textbf{Flux relaxation} $\Gamma_{1,{\Phi_x}}\propto|f_{01}|$\\
\vspace*{-2mm}\\
Being $\partial_{\Phi_x} \hat{U}\sim(\cos\hat\varphi+\cos2\hat\varphi)$ a $\hat{K}$-symmetric operator, the matrix element $f_{01}$ vanishes in the twist-based regime and in the plasmonic one for $\langle n_g\rangle=0$. In the flux-biased regime, for $E_C\ll E_J$, it is also suppressed due to the localization of the wavefunctions around $\vf=0$ and $\vf=\pi$ as shown in Fig.~\ref{FS4}(d). More specifically, in this regime, the dominant term in $f_{01}$ is $\cos2\hat\vf$. The contribution of $\cos\hat\vf$ almost cancels out  since the two wavefunctions are substantially equal up to a $\pi$-translation and they are both $\hat K$-symmetric, while $\cos\hat\vf$ changes sign at $\vf=\pi/2$. 
Analogously, $f_{01}$ stays exponentially suppressed for $E_C\ll E_J$ and $\langle n_g\rangle\neq0$, see Fig.~\ref{FS5}(d). A significant increase of $f_{01}$ is instead found for asymmetric junctions even at $\langle n_g\rangle=0$. Nevertheless we note that $f_{01}$ remains  at least one order of magnitude smaller than $f_{11}-f_{00}$, see Fig.~\ref{FS6}.
 \begin{figure}[H]
\hspace*{20mm}
    \begin{minipage}{0.45\textwidth}\hspace*{-9mm}\includegraphics[scale=0.55]{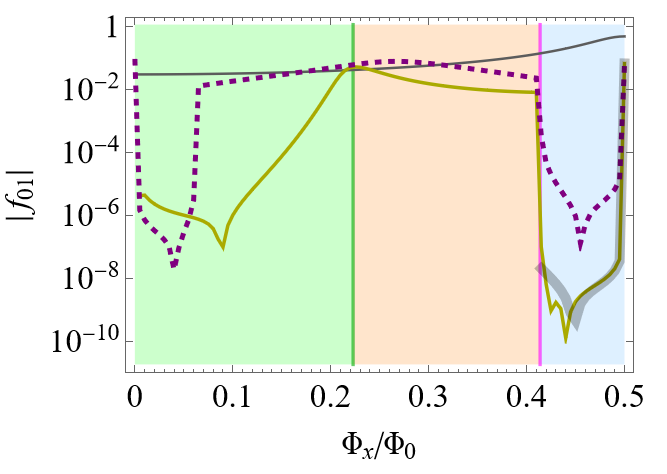}
    \end{minipage}
  \begin{minipage}{0.4\textwidth}
  \caption{The matrix element $|f_{01}|$ governing flux relaxation for $\langle n_g\rangle=0$ and  $d=d_{\kappa}=10\%$ (solid yellow curve), $d=d_{\kappa}=10\%,\,d_{\theta}=5\% $ (dashed purple curve) compared to the case $d=d_\theta=d_\kappa=0$ (shaded thick gray curve) and to the standard asymmetric transmon with $d=10\%$ (thin gray curve). The parameters $d$, $d_\kappa$ and $d_\theta$ are defined as in the main text.}
    \label{FS6}
  \end{minipage}
\end{figure}
\end{itemize}

\section{Supersymmetry}\label{SUSY}
\noindent For a specific value $\phi_x^{\scalebox{0.7}{\rm SUSY}}$ of the external flux  all the excited levels of the Hamiltonian \eqref{e. Hamiltonian f-SQUID} become degenerate due to the emergence of a supersymmetry. In this section we give an analitic expression of $\phi_x^{\scalebox{0.7}{\rm SUSY}}$ which, as pointed out in Sec.~\ref{diff regimes}, is the threshold value separating the plasmonic regime from the flux-biased one. 
\begin{figure}[H]
   \centering
   \includegraphics[scale=0.6]{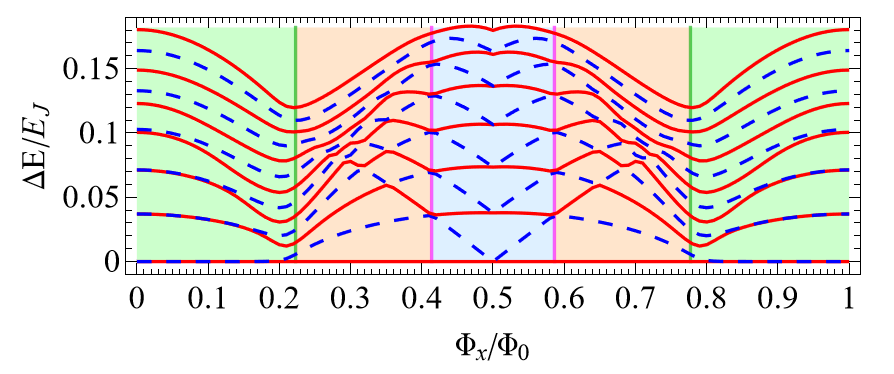}
   \caption{Spectrum for $\langle n_g\rangle=0.25$ showing that the degeneracy associated with supersymmetry is lifted in the high-energy levels.}
    \label{FS7}
\end{figure}
\noindent A given Hamiltonian $\hat H$ is said to be supersymmetric if there exists a Hermitian supercharge $\hat Q$ such that $\hat H=\hat Q^2+c$ with $c\in\mathbb{R}$, and if there exists a Hermitian involution $\hat K$, with $\hat{K}^2=1$, that anticommutes with $\hat Q$, $\{\hat K,\hat Q\}=0$. It then follows that the involution $\hat K$ commutes with the Hamiltonian and that the spectrum can be split in the direct sum of two sectors, each one having a definite eigenvalue of $\hat K$, so that $\hat{H}=\hat{P}_+\hat{H}\hat{P}_+ + \hat{P}_-\hat{H}\hat{P}_-$, with $\hat{P}_\pm=(1\pm \hat{K})/2$. The two sectors are exchanged through $\hat{P}_\pm \hat{Q}=\hat{Q}\hat{P}_\mp$ in a way that, for every state $\ket{\psi_i}$ of $\hat{H}$ with eigenvalue $\hbar\omega_i$, $\hat{H}\ket{\psi_i}=\hbar \omega_i\ket{\psi_i}$, there exists a supersymmetric partner $\hat{Q}\ket{\psi_i}$ at the same energy. 
To be specific, let us consider the Hamiltonian describing the device presented in this work for $\langle n_g\rangle=0$,  
\begin{equation}\label{EqSupMat-Ham}
\hat H=4E_C\hat{n}^2-E_J\cos(2\theta)\cos(\phi_x)\cos(\hat\varphi)+E_\kappa\cos(2\phi_x)\cos(2\hat\varphi)~.
\end{equation}
Following the route outlined in Refs.~\cite{ulrich2014,ulrich2015} it is possible to show that this Hamiltonian can be written as $\hat H\sim \hat{Q}^2$, defining $\hat Q$ as 
\begin{equation}
\hat Q=2\sqrt{E_C}
\left(\hat n+i\sqrt{\frac{-E_\kappa\cos(2\phi_x)}{2E_C}}\sin(\hat\varphi)\right)(-1)^{\hat n}~.
\end{equation}
and fixing the flux so that the following condition is satisfied
\begin{equation}\label{fixsusy}
    E_J\cos(2\theta)\cos(\phi_x)=\sqrt{-8E_CE_\kappa \cos(2\phi_x)}~.
\end{equation}
It is important to note that the operator $\hat {Q}$ is Hermitian only if $\sqrt{-E_\kappa\cos(2\phi_x)}$ is real, therefore, independently on the value of $E_C$ and $E_\kappa$, for Eq.~\ref{fixsusy} to hold we must have $\pi/4<\phi_x<\pi/2$. It then follows that 
\begin{equation}
    \phi_x^{\scalebox{0.7}{\rm SUSY}}=\frac{1}{2}\arccos{\left(\frac{-1}{1+16E_\kappa  E_C/(E^2_J\cos^2(2\theta))}\right)}~.
\end{equation}
In particular for $\theta=43^\circ$, $\alpha=0.1$ and $E_J/E_C=2000$ as in the main text, $\phi_x^{\scalebox{0.7}{\rm SUSY}}\sim 0.414\,\pi$ in accordance with Fig.~\ref{FS4}. It is worth noting that supersymmetry guarantees doubly degeneracy of the spectrum a part from the ground state. The operator $\hat{Q}$ is Hermitian and therefore eigenstates of $\hat Q$ with eigenvalue $q_i$, $\hat Q\ket{\psi_i}=q_i\ket{\psi_i}$, are also eigenstates of the Hamiltonian with energy $\hbar \omega_i=q_i^2$ a part from an overall shift of the zero energy. It follows that a state for which $\hat{Q}\ket{\psi_0}=0$ is the ground state. Its non degeneracy comes from the fact that it can only belong to the $\hat{K}$-symmetric sector. As shown by Fig.~\ref{FS7}, if $\langle n_g\rangle\neq 0$ supersymmetry is broken, however the quasi-degeneracy of the low energy states is preserved.

\section{$\pm\pi/2$ qubit and $0-\pi$ qubit}\label{0pi_vs_pihalf}

\begin{figure}[H]
   \centering
   \includegraphics[scale=0.55]{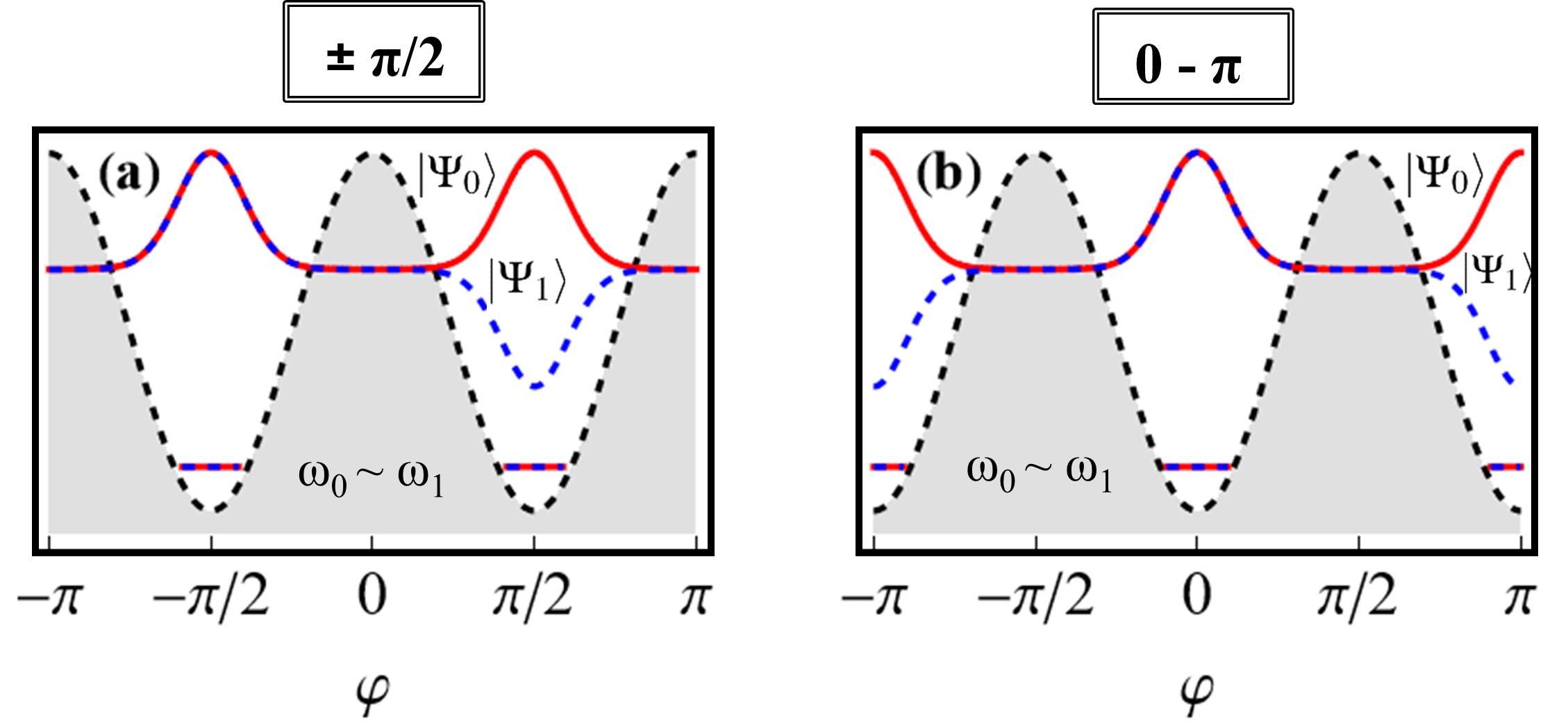}
   \caption{First two levels and wavefunctions for \textbf{(a)} $\pm\pi/2$ and \textbf{(b)} $0-\pi$ qubit. The solid red lines identify $\omega_0$ and $\ket{\Psi_0}$, while the dashed blue ones  $\omega_1$ and $\ket{\Psi_1}$. }
    \label{FS8}
\end{figure}

\noindent Here we consider the structure of the spectrum and the symmetry of the eigenstates at $\theta\sim 45^\circ,\, \phi_x=0$ and at $\phi_x=\pi/2$. 
In both these situations the first harmonic Josephson tunneling vanishes while the second remains finite and dominates over the charging energy
so that the circuit implements, respectively, 
a $\pm\pi/2$ and a $0-\pi$ qubit. The $\pm\pi/2$  Hamiltonian can be cast as 
\begin{equation}
\hat H_{\pm\pi/2}=4E_C\hat n^2+E_\kappa\cos(2\hat\varphi)~,
\end{equation}
while the $0-\pi$ Hamiltonian reads
\begin{equation}
\hat H_{0-\pi}=4E_C\hat n^2-E_\kappa\cos(2\hat\varphi)~.
\end{equation}
The $\pi$-periodicity of the potential implies that the Hamiltonian commutes with the Hermitian operator $\hat\Pi=(-1)^{\hat n}=e^{i\pi \hat{n}}$. This property yields a separation of the Hilbert space in eigenstates that are composed by superposition of  even charge states, symmetric under $\hat\Pi$, and eigenstates featuring only odd charge states, antisymmetric under $\hat\Pi$. In addition, the invariance of the potential under $\hat K$ forces the eigenstates to have a definite $\hat K$-parity. Summarizing for both Hamiltonians we have the following symmetries:
\be
 \hat\Pi=(-1)^{\hat n}=e^{i\pi \hat{n}} \qquad{\rm and }\qquad
 \hat K\;\mbox{ such that }\;\hat K\hat\vf\hat K=-\hat\vf~.
\ee
In both cases we can therefore label every eigenstate by a principal quantum number $m$, the parity $s_p$ under $\hat \Pi$, and the parity $s_k$ under $\hat K$, so that
\be \hat\Pi \ket{m,s_p,s_k}=s_p\ket{m,s_p,s_k}~,\quad   \hat K\ket{m,s_p,s_k}=s_k\ket{m,s_p,s_k}~,\;\; \ee
with $s_p,s_k\in\{-1,+1\}$.
These properties hold true for both the $\pm \pi/2$ and the $0-\pi$ Hamiltonians hinting at a one-to-one correspondence between their eigenstates.
The two Hamiltonians are indeed connected by a unitary transformation $\hat{U}=e^{i\pi \hat{n}/2}$ such that $\hat{H}_{0-\pi}=\hat{U}^\dag \hat{H}_{\pm \pi/2}\hat{U}$.
We thus have that, given an eigenstate $\ket{m,s_p,s_k}$ of $\hat{H}_{0-\pi}$ with eigenvalue $\omega_m$, {\it i.e.} $\hat{H}_{0-\pi}\ket{m,s_p,s_k}=\hbar \omega_m\ket{m,s_p,s_k}$, the state $\hat{U}\ket{m,s_p,s_k}$ is eigenstate of $\hat{H}_{\pm\pi/2}$ with the same eigenvalue $\omega_m$, {\it i.e.} $\hat{H}_{\pm\pi/2}\hat{U}\ket{m,s_p,s_k}=\hbar\omega_m\hat{U}\ket{m,s_p,s_k}$. In addition, we have that $\hat{K}\hat{U}=\hat{U}^\dag \hat{K}$ and $\hat{U}^2=\hat\Pi$. It follows that, given an eigenstate  $\ket{m,s_p,s_k}$ of $\hat{H}_{0-\pi}$  we have 
\begin{eqnarray}
\hat{K}\hat{U}\ket{m,s_p,s_k}&=&\hat{U}^\dag \hat{K} \ket{m,s_p,s_k}\nonumber\\
&=& s_k\hat{U}^\dag  \ket{m,s_p,s_k}\nonumber\\
&=& s_k \hat{U}^3\ket{m,s_p,s_k}\nonumber\\
&=&s_k s_p\hat{U}\ket{m,s_p,s_k} 
\end{eqnarray}
where we used  $(\hat{U}^\dag)^4=\hat{U}^4=1$. 
Eventually,  using  $\hat{U}^2=\hat{\Pi}$, we obtain
\be \hat{U}\ket{m,s_p,s_k}=\ket{m,s_p,s_k s_p}~.\label{e. 0pi to pmpihalf}\ee
Finally in Fig.~\ref{FS8} we plot the wavefunctions $\ket{\Psi_0}$ and $\ket{\Psi_1}$ together with the potential in both cases.

\bibliography{Biblio_sup}